\documentclass[preprint,showpacs,preprintnumbers,amsmath,amssymb]{revtex4}
\usepackage{graphicx}
\usepackage{dcolumn}
\usepackage{bm}

\begin{document}


\title{Coherent Dissociation $^{16}$O~$\rightarrow$~4$\alpha$ in Photoemulsion at an Incident Momentum of 4.5 GeV/$c$ per Nucleon}

\author{N.~P.~Andreeva}
   \affiliation{Institute of High Energy Physics. National Academy of Sciences of Kazakhstan, pos. Ala-Tau, Almaty, 480082 Kazakhstan}  
\author{Z.~V.~Anzon}
   \affiliation{Institute of High Energy Physics. National Academy of Sciences of Kazakhstan, pos. Ala-Tau, Almaty, 480082 Kazakhstan}
 \author{V.~I.~Bubnov}
   \affiliation{Institute of High Energy Physics. National Academy of Sciences of Kazakhstan, pos. Ala-Tau, Almaty, 480082 Kazakhstan}
\author{A.~Sh.~Gaitinov}
   \affiliation{Institute of High Energy Physics. National Academy of Sciences of Kazakhstan, pos. Ala-Tau, Almaty, 480082 Kazakhstan}
\author{G.~Zh.~Eligbaeva}
   \affiliation{Institute of High Energy Physics. National Academy of Sciences of Kazakhstan, pos. Ala-Tau, Almaty, 480082 Kazakhstan}
\author{L.~E.~Eremenko}
   \affiliation{Institute of High Energy Physics. National Academy of Sciences of Kazakhstan, pos. Ala-Tau, Almaty, 480082 Kazakhstan}
\author{G.~S.~Kalyachkina}
   \affiliation{Institute of High Energy Physics. National Academy of Sciences of Kazakhstan, pos. Ala-Tau, Almaty, 480082 Kazakhstan}
\author{E.~K.~Kanygina}
   \affiliation{Institute of High Energy Physics. National Academy of Sciences of Kazakhstan, pos. Ala-Tau, Almaty, 480082 Kazakhstan}
\author{A.~M.~Seitimbetov}
   \affiliation{Institute of High Energy Physics. National Academy of Sciences of Kazakhstan, pos. Ala-Tau, Almaty, 480082 Kazakhstan}
\author{I.~Ya.~Chasnikov}
   \affiliation{Institute of High Energy Physics. National Academy of Sciences of Kazakhstan, pos. Ala-Tau, Almaty, 480082 Kazakhstan}
\author{Ts.~I.~Shakhova}
   \affiliation{Institute of High Energy Physics. National Academy of Sciences of Kazakhstan, pos. Ala-Tau, Almaty, 480082 Kazakhstan}

\author{M.~Haiduk}
   \affiliation{Institute of Physics, Romania Academy of Sciences, Bucharest. Romania}

\author{S.~A.~Krasnov}
   \affiliation{Joint Insitute for Nuclear Research, Dubna, Russia, Moscow oblast', 141980 Russia}
\author{T.~N.~Maksimkina}
   \affiliation{Joint Insitute for Nuclear Research, Dubna, Russia, Moscow oblast', 141980 Russia}
\author{K.~D.~Tolstov}
   \affiliation{Joint Insitute for Nuclear Research, Dubna, Russia, Moscow oblast', 141980 Russia}
\author{G.~M.~Chernov}
   \affiliation{Joint Insitute for Nuclear Research, Dubna, Russia, Moscow oblast', 141980 Russia}

\author{N.~A.~Salmanova}
   \affiliation{Tadzhik State University, Dushanbe. Tadzhikistan}
\author{D.~A.~Salomov}
   \affiliation{Tadzhik State University, Dushanbe. Tadzhikistan}
\author{A.~Khushvaktova}
   \affiliation{Tadzhik State University, Dushanbe. Tadzhikistan}
 
\author{F.~A.~Avetyan}
   \affiliation{Yerevan Physics Institute, ul. Brat'ev Alikhanian 2. Yerevan, 375036 Armenia}
\author{N.~A.~Marutyan}
   \affiliation{Yerevan Physics Institute, ul. Brat'ev Alikhanian 2. Yerevan, 375036 Armenia}
\author{L.~T.~Sarkisova}
   \affiliation{Yerevan Physics Institute, ul. Brat'ev Alikhanian 2. Yerevan, 375036 Armenia}
\author{V.~F.~Sarkisyan}
   \affiliation{Yerevan Physics Institute, ul. Brat'ev Alikhanian 2. Yerevan, 375036 Armenia}

\author{M.~I.~Adamovich}
   \affiliation{Lebedev Institute of Physics, Russian Academy of Sciences. Leninskii pr. 53. Moscow, 117924 Russia}
\author{Yu.~A.~Bashmakov}
   \affiliation{Lebedev Institute of Physics, Russian Academy of Sciences. Leninskii pr. 53. Moscow, 117924 Russia}
\author{V.~G.~Larionova}
   \affiliation{Lebedev Institute of Physics, Russian Academy of Sciences. Leninskii pr. 53. Moscow, 117924 Russia}
\author{G.~I.~Orlova}
   \affiliation{Lebedev Institute of Physics, Russian Academy of Sciences. Leninskii pr. 53. Moscow, 117924 Russia}
\author{N.~G.~Peresadko}
   \affiliation{Lebedev Institute of Physics, Russian Academy of Sciences. Leninskii pr. 53. Moscow, 117924 Russia}
\author{M.~I.~Tretyakova}
   \affiliation{Lebedev Institute of Physics, Russian Academy of Sciences. Leninskii pr. 53. Moscow, 117924 Russia}
\author{S.~P.~Kharlamov}
   \affiliation{Lebedev Institute of Physics, Russian Academy of Sciences. Leninskii pr. 53. Moscow, 117924 Russia}
\author{M.~M.~Chemyavsky}
   \affiliation{Lebedev Institute of Physics, Russian Academy of Sciences. Leninskii pr. 53. Moscow, 117924 Russia}

\author{V.~G.~Bogdanov}
   \affiliation{Khopin Radium Institute, St. Petersburg, 197022 Russia}
\author{V.~A.~Plyushchev}
   \affiliation{Khopin Radium Institute, St. Petersburg, 197022 Russia}
\author{Z.~I.~Soloveva}
   \affiliation{Khopin Radium Institute, St. Petersburg, 197022 Russia}
   
\author{V.~V.~Belaga}
   \affiliation{Institute of Nuclear Physics. Academy of Sciences of Uzbekistan, pos. Ulughbek, Tashkent, 702132 Uzbekistan}
\author{A.~I.~Bondarenko}
   \affiliation{Institute of Nuclear Physics. Academy of Sciences of Uzbekistan, pos. Ulughbek, Tashkent, 702132 Uzbekistan}
\author{Sh.~A.~Rustamova}
   \affiliation{Institute of Nuclear Physics. Academy of Sciences of Uzbekistan, pos. Ulughbek, Tashkent, 702132 Uzbekistan}
\author{A.~G.~Chemov}
   \affiliation{Institute of Nuclear Physics. Academy of Sciences of Uzbekistan, pos. Ulughbek, Tashkent, 702132 Uzbekistan}

\author{N.~N.~Kostanashvili}
   \affiliation{Tbilisi State University, Universitetskaya ul. 9, Tbilisi, 380086 Georgia}

\author{S.~Vok\'al}
   \affiliation{P. J. \u Saf\u arik University, Ko\u sice, Jesenna 5, SK-04154, Slovak Republic}
   
\date{\today}

\begin{abstract}
\indent 
First searches for the coherent dissociation of relativistic oxygen nuclei into four a particles are reported. It is shown that reactions of this type are characterized by a significantly lower decay temperature than the conventional multifragmentation of residual projectile nuclei. The momentum spectra and correlations of a panicles are not reproduced by the simple statistical model of direct fragmentation. The possibility that the oxygen nucleus undergoing fragmentation acquires a nonzero angular momentum in the collision process is discussed.\par
\indent \par
\indent DOI: 1063-7788/96/5901-0102S10.00\par
\end{abstract}
 \pacs{21.45.+v,~23.60+e,~25.10.+s}

\maketitle

\indent Multifragmentation of relativistic projectile nuclei provides unique insight into their structure: as charged secondaries are detected without any thresholds, the corresponding transitions can be observed at the lowest possible values of 4-momentum transfer. Extremely peripheral, coherent, inelastic nuclear collisions in which the target nucleus acts on the dissociating projectile nucleus as a discrete particle - that is, the former does not change charge, does not undergo breakup, and is not excited - are especially promising in this respect. Investigations of such collisions between high-energy particles were initiated largely by the milestone theoretical results by Pomeranchuk and Feinberg \cite{Pomeranchuk}. Note, however, that these results equally apply to collisions in which high-energy nuclei take the place of particles as projectiles, while the products of nuclear fragmentation appear as final states instead of elementary particles produced \cite{Chernov}.\par

\indent As in coherent particle production on nuclei, processes of two types - a diffractive transition mediated by a Pomeron and a Coulomb transition mediated by a virtual photon - occur in the above nuclear reactions. The latter is expected to be dominant for colliding nuclei with high electric charges. Typically, inelastic coherent reactions have rather high energy thresholds. For example, the threshold incident momentum for coherent dissociation of a projectile nucleus $A$ with mass $M_0$ into $n$ fragments with masses $m_i(i~=~1,...,n)$ is estimated as \cite{Chernov} 
$$
p^{min}_0~\approx~(M_0B^{1/3}/\mu)\Delta~~~~~(1)
$$ 
where $\mu$ is the pion mass, $B$ is the mass number of the target nucleus that coherently interacts with the projectile $A$, and $\Delta~=~\sum^n_{i=1}m_i~-~M_0$ is the mass defect with respect to the dissociation channel considered. In particular, for the reaction $^{16}$O~$\rightarrow$~4$\alpha$, the momentum threshold amounts to several hundred MeV/$c$ per nucleon, approaching the relativistic region for heavy targets.\par

\indent Only a few studies devoted to the coherent dissociation of relativistic projectile nuclei have been performed thus far. This paper reports the results of first searches for the coherent dissociation process 
$$
^{16}O~\stackrel{Em}{\longrightarrow}~4\alpha~~~~~(2)
$$
at $p_0$~=~4.5~GeV/$c$ per nucleon in nuclear emulsion (Em). Earlier, the dissociation of oxygen nuclei into four $\alpha$ particles was investigated at nonrelativistic energies (below 90 MeV per nucleon; see \cite{Pouliot,Badala} and references therein), which probably fall short of the coherent threshold at least for the heavy target nuclei.\par

\indent Stacks of nuclear emulsion BR-2 irradiated by a beam of $^{16}$O ions with a momentum of 4.5~GeV/$c$ per nucleon from the JINR synchrophasotron were used in searches for events satisfying criteria necessary to select reaction~(2).\par

\indent To estimate the mean range $\lambda$ of $^{16}$O nuclei in emulsion for reaction~(2) without additional charged secondaries, a slow-fast scan along the track was carried out in a part of emulsion sheets. Over the primary track of total length 375.2~m, we found 12 pure events of reaction~(2). This corresponds to a $\lambda$ value of 31.3$^{+12.6}_{-7.0}$~m. From this, the cross section for the so-called average emulsion nucleus $(\langle A\rangle\approx45)$ was estimated at 10~mb/nucleus.\par 	

\indent To maximize the statistics of events we are interested in, we made use of a special procedure based on a fast area scan in a large number of emulsion sheets. The scan was performed over bands normal to the incident oxygen beam. The distance between the bands was 0.5~cm. In the scan, we selected close groups comprising 3 to 4 gray tracks of similar ionization. Once such a group has been located, search along the group axis was carried out in the reverse direction. In comparison with scanning along the track, this strategy resulted in more than a tenfold increase in the scanning efficiency for reaction~(2).\par

\indent In this way, we found and measured as many as 641 events that involve four well-identified relativistic fragments with $z$~=~2 in the final state and which show evidence neither for the emergence of additional charged particles, nor for the excitation of the target nucleus. These events are further analyzed in the present study.\par

\begin{figure}
    \includegraphics[width=4in]{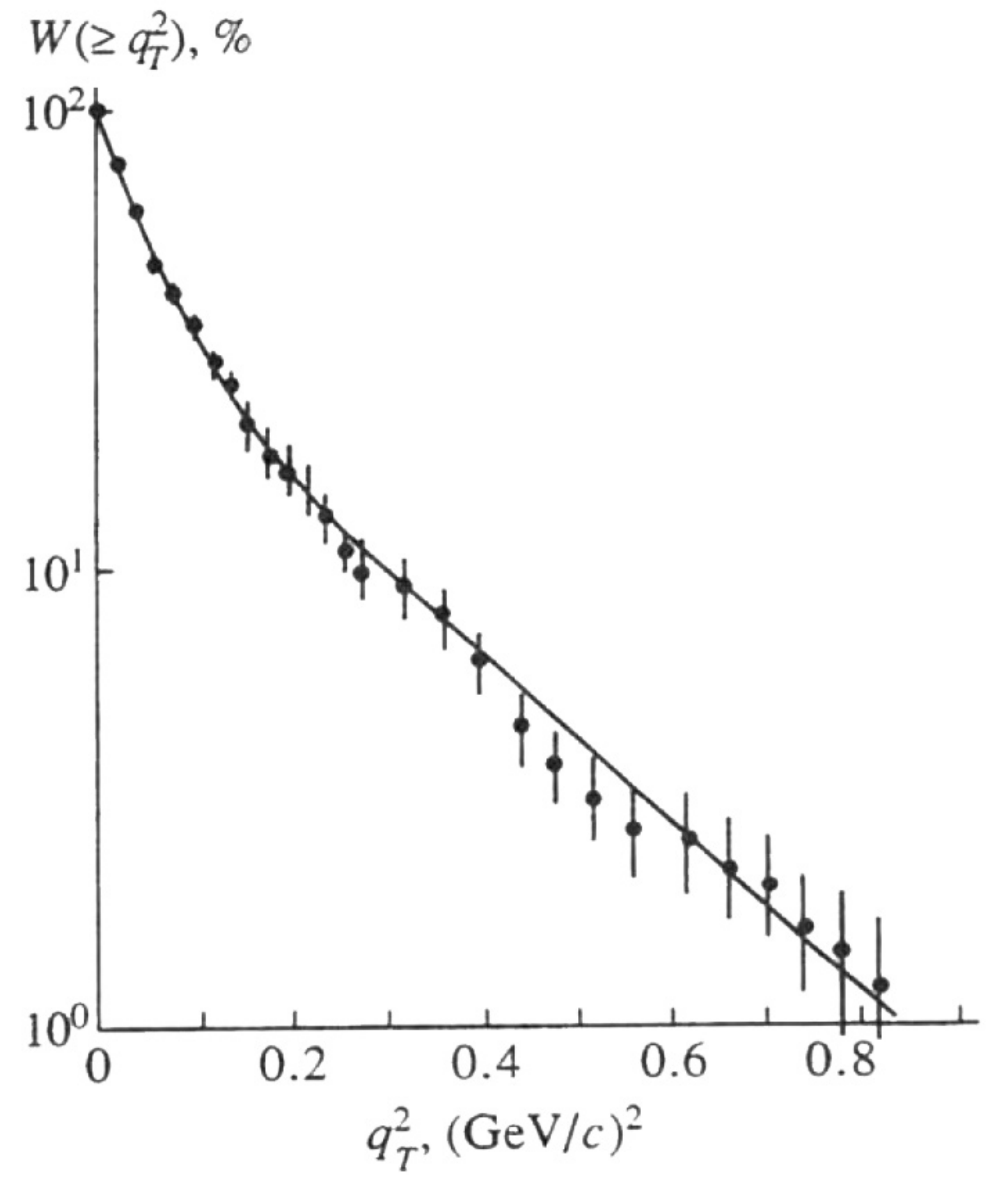}
    \caption{\label{Fig:1} Distribution of $^{16}$O~$\rightarrow$~4$\alpha$ events in $q^2_T=-t'$. A fit to the functional form~(6) is shown by a curve.}
    \end{figure}
		
\indent Figure 1 shows the distribution in the square $q^2_T$ of the transverse-momentum transfer to the $^{16}$O nucleus through interaction with an emulsion nucleus. This quantity can be used to estimate the square $t$ of the 4-momentum transfer to the oxygen nucleus splitting into $n$ fragments. This is done with the aid of the relation
$$
|t'|~=~|t-t^{min}(\sum_{i=1}^{n}m_i)|~\simeq~q^2_T.~~~~~(3)
$$
Indeed, in the coherent diffractive production of the fragments (as in elastic diffractive scattering), the distribution in $t'$ {and in $t$ as well, because the factor exp$[t^{min}(\sum_{i=1}^nm_i)]$ that takes into account the minimum 4-momentum transfer squared $t^{min}(\sum_{i=1}^nm_i)$ required for the production of $n$ free fragments is insignificant here} must have the simple exponential form
$$
d\sigma/dt'~\varpropto~exp(-a|t'|),~~~~~(4)
$$
where the slope of the diffractive peak $a$ is expressed in terms of the radii of the projectile and target nuclei, $R_A$ and $R_B$, as $a~\approx~(R_A~+~R_B)^2/4$. On average, the resulting longitudinal component of the 3-momentum transfer $\mathbf q$ is much less than the transverse component. It follows that $-t'~\approx~q^2_T$; hence, the distribution in the transverse momentum squared must have the Rayleigh form
$$
d\sigma/dq^2_T~\varpropto~exp(-aq^2_T)~~~~~(5)
$$
with $\langle q_T\rangle~\approx~\pi^{1/2}/(R_A~+~R_B)$. We assumed that there are no neutral fragments in the events under study. In this case, we have $\mathbf q_T~=~\sum_{i=1}^4\mathbf p_T$, where $\mathbf p_T$ is the $\alpha$-particle transverse momentum. Its absolute value $p_T$ was estimated by the formula $p_T~=~4p_0sin\theta$, where $p_0~=~4.5$~GeV/$c$ and $\theta$ is the polar emission angle. A straight line corresponds to distribution (5) on the exponential scale used in Fig.~1.\par

\indent It can be seen from Fig. 1 that the experimental distribution in $q^2_T$ is not consistent with the functional form (5) ($\chi^2$/NDF~=~3.2). However, it can be described by the sum of two Rayleigh distributions,
$$
d\sigma/dq^2_T~=~\alpha exp(-a_1q^2_T)~+~(1~-~\alpha)exp(-a_2q^2_T),~~~~~(6)
$$
with two different slope parameters $a_1$ and $a_2$. A maximum-likelihood fit of (6) to experimental data yields ($\chi^2$/NDF~=~0.9)
$$
a_1~=~19~\pm~2,~a_2~=~4.2~\pm~0.4,~\alpha~=~0.66~\pm~0.06.~~~~~(7)
$$
Once the nuclear composition of emulsion and experimental uncertainties in $q_T$, which effectively broaden the distribution in $q_T$, are taken into account, the resulting values of $a_1$ and $a_2$ are compatible with the assumption that the first and second terms in (6) correspond, respectively, to the diffractive dissociation $^{16}$O~$\rightarrow$~4$\alpha$ on an emulsion nucleus as a discrete particle and dissociation on a nucleon. At the same, any attempts at estimating the relative contributions of these two dissociation channels (that is, the slope parameters $\alpha$) and their cross sections are seriously impeded by the fact that events with a recoil proton are not observed in our experiment, by the possible contribution of Coulomb dissociation, by the area scan used, etc.\par

\indent Along with the complete sample of 641 pure events of the type $^{16}$O~$\rightarrow$~4$\alpha$, we will henceforth consider the subsample of 428 ($\sim67\%$ of the complete sample) events that satisfy the condition $q_T~<~0.32$~GeV/$c$, which is equivalent to $葉'~<~0.1~($GeV/$c)^2$. This value of $t'$ (see Fig. 1) may be taken for an approximate boundary between the coherent and incoherent events in the selected sample. Provided that the coherent reaction (2) is indeed responsible for the first term in (6), this separation enables us to constrain from below the characteristics of the two reactions. In other words, we may hope to obtain some reliable estimates by comparing the characteristics of the complete sample and of the subsample with $-t'~<~0.1$~(GeV/$c)^2$.\par

\begin{figure}
    \includegraphics[width=3in]{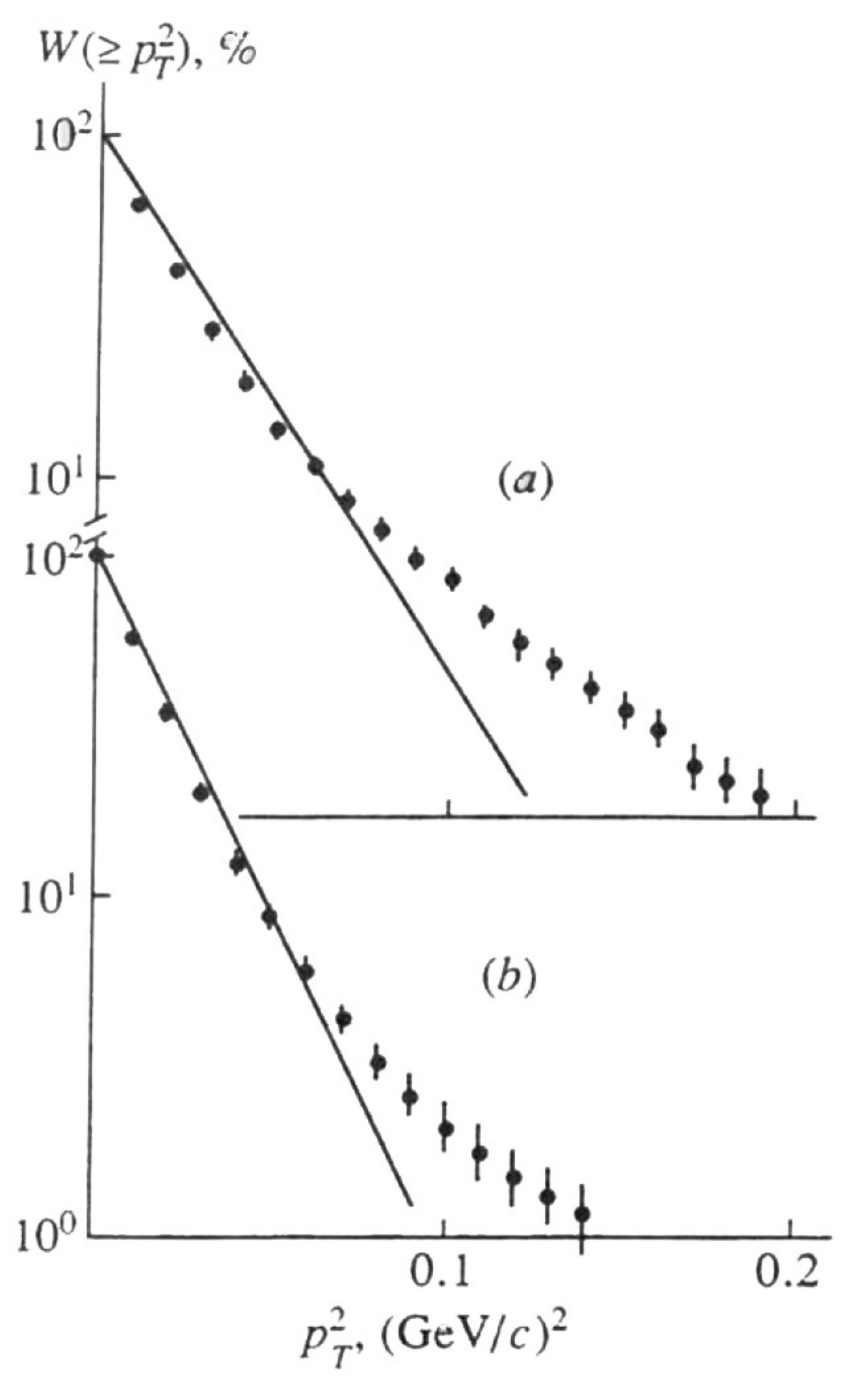}
    \caption{\label{Fig:2} Inclusive $p^2_T$ distribution of $\alpha$ particles for ($a$) the complete event sample and ($b$) events with $-t'~<~0.1$~(GeV/$c$). The straight lines are the Rayleigh distributions with $\langle p^2_T\rangle$ corresponding to the observed values.}
\end{figure}
	
\indent Figure 2 shows the transverse-momentum-squared distributions of $\alpha$ particles from reaction (2) for the complete sample and for the subsample with $-t'~<~0.1$~(GeV/$c)^2$. The corresponding root-mean-square values $\langle p^2_T\rangle^{1/2}$ are 167~$\pm$~4 and 145~$\pm$~4~MeV/$c$. The two distributions display deviations from the Rayleigh form, showing high-energy tails that are due to particles with high $p_T$. These deviations cannot be explained by the complex composition of emulsion, in view of the well-known result from mathematical statistics that an arbitrary superposition of any number of Rayleigh distributions with different a values is a Rayleigh distribution as well.\par

\begin{figure}
    \includegraphics[width=3in]{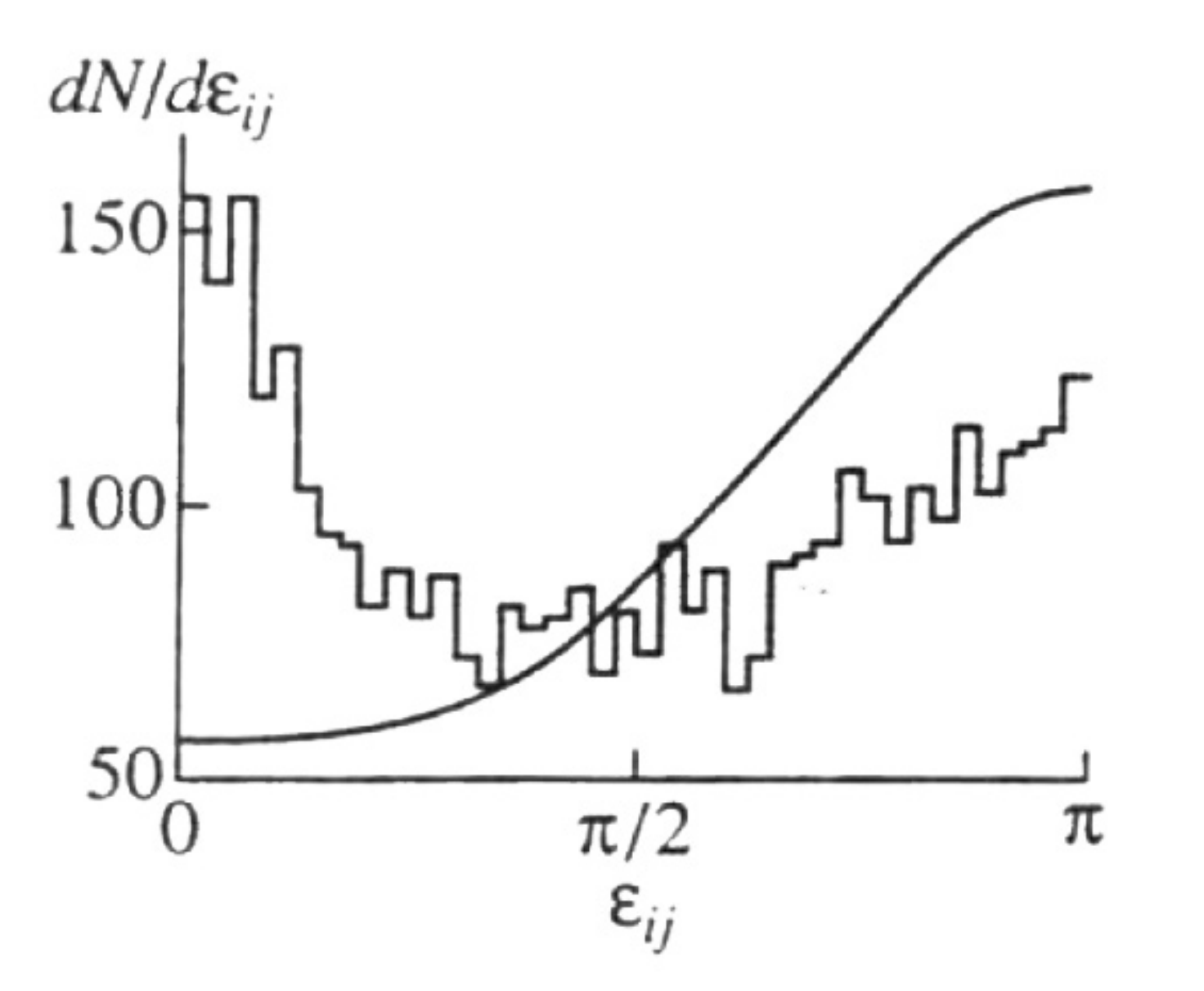}
    \caption{\label{Fig:3} Distribution in the $\alpha\alpha$ azimuthal angle $\varepsilon_{ij}$ for the complete sample of $^{16}$O~$\rightarrow$~4$\alpha$ events.}
    \end{figure}
	
\indent It is also well known that the laboratory transverse momenta of fragments are effectively overestimated because of the transverse motion of the nucleus underｬgoing fragmentation (the so-called bounce-off effect discussed, for example, in \cite{Bengus}). In our data, this effect manifests itself in the distribution in the azimuthal angle $\varepsilon_{ij}$~=~arccos($\mathbf p_{T_i}\mathbf p_{T_j}/p_{T_i}p_{T_j}$) between the transverse
momenta of two $\alpha$ particles produced in the same event of dissociation (2) (see Fig. 3). This distribution contradicts the phase-space prediction (the curve in the same figure). In particular, for the observed azimuthal asymmetry
$$
A~=~(N_{\varepsilon_{ij}\ge\pi/2}~-~N_{\varepsilon_{ij}<\pi/2})/N_{0\le\varepsilon_{ij}\le\pi}~~~~~(8)
$$
of the distribution $d\sigma/d\varepsilon_{ij}$, we have -0.01~$\pm$~0.02, while the value predicted by the law of transverse-momentum conservation is $1/(N_{\alpha}$~-~1)~=~0.33~\cite{Bondarenko}. This means that the fragmentation of the oxygen nucleus must be described in terms of the quantities defined in its rest frame.\par

\indent For an exclusive reaction like (2), the recipe for going over to the rest frame of the nucleus undergoing fragmentation is straightforward. (In the following, all the quantities that refer to this frame are labeled with asterisks.) Under the assumption that the projectile nucleus is scattered at a small angle, a high-precision approximation to the transverse momenta of $\alpha$ particles is given by the formula
$$
\mathbf p^*_{T_i}~\cong~\mathbf p_{T_i}~-~\sum_{i=1}^4\mathbf p_{T_i}/4~~~~~(9)
$$
Figures 4 and 5 show, respectively, the $\mathbf p^*_{T_i}$ and $\varepsilon^*_{ij}$ distributions of $\alpha$ particles for ($a$) the complete sample of events (2) and for ($b$) the subsample with $葉'~<~0.1$~(GeV/$c$). The main quantitative characteristics of these distributions are given in the table. These are the mean values $\langle p^*_T\rangle$ and $\langle p^{*2}_T\rangle^{1/2}$, the azimuthal asymmetry $A^*$ [see formula (8) for $\varepsilon_{ij}$], and the azimuthal collinearity $B^*$ defined as
$$
B^*~=~(N_{\varepsilon^*_{ij}\le\pi/4}~+~N_{\varepsilon^*_{ij}\ge3\pi/4}~-~N_{\pi/4<\varepsilon^*_{ij}<3\pi/4})/N_{0\le\varepsilon^*_{ij}\le\pi}.~~~~~(10)
$$
The data presented in Figs. 4 and 5 and in the table lead to the following conclusions.
\par

\begin{figure}
    \includegraphics[width=3in]{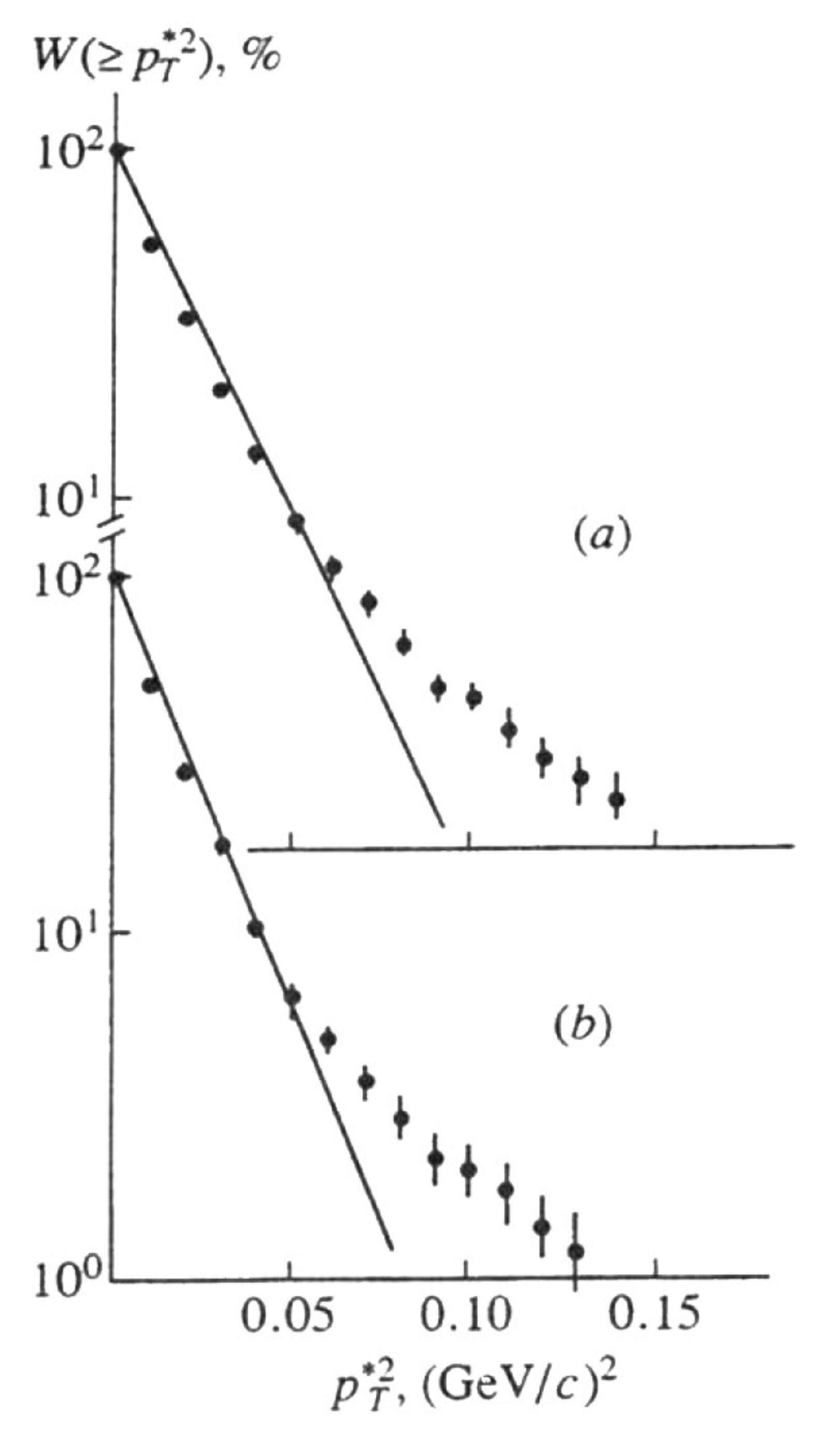}
    \caption{\label{Fig:4} As in Fig. 2, but for the $\alpha$ transverse momentum defined in the $^{16}$O rest frame.}
\end{figure}

\begin{figure}
    \includegraphics[width=3in]{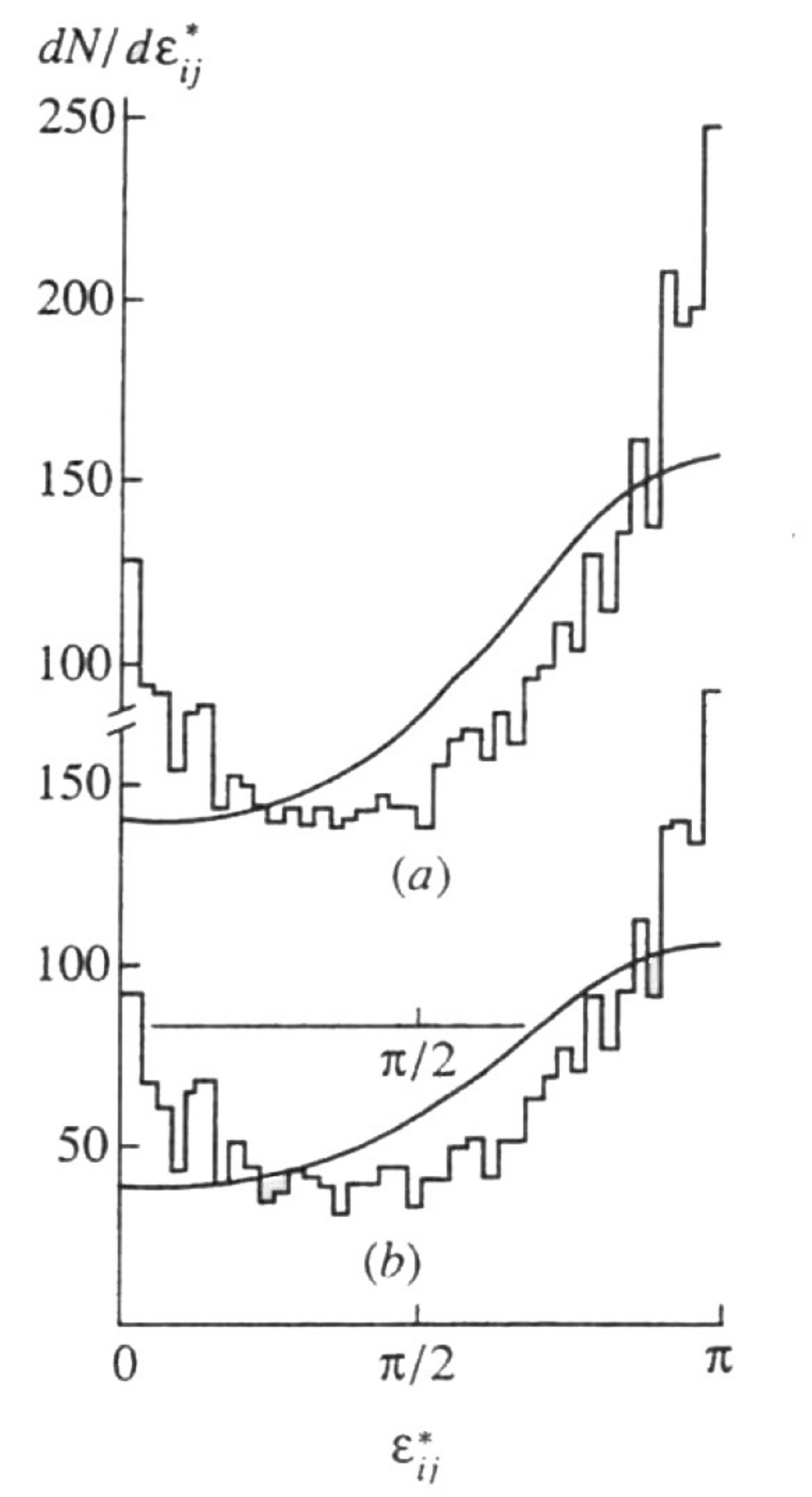}
    \caption{\label{Fig:5} Distributions in $\varepsilon^*_{ij}$ for ($a$) the complete event sample and ($b$) events with $-t'~<~0.1$~(GeV/$c)^2$ The curves represent distribution (11) with coefficients~(12).}
\end{figure}
	
\indent (1)~As might have been expected, the mean values $\langle p^*_T\rangle$ are substantially smaller than $\langle p_T\rangle$ (by some 30\% for the complete sample of $^{16}$O~$\rightarrow$~4$\alpha$). It is noteworthy that the $\langle p^*_T\rangle$ values for all events and for those with $-t'~<~0.1$~(GeV/$c)^2$ differ only slightly.\par

\indent (2)~As in the laboratory frame, the $p^{*2}_T$ distributions show deviations from the Rayleigh form for both sets. It should be emphasized that within errors, the relative contribution of the high-$p^*_T$ tail for the complete sample does not differ from that for the subsample with $-t'~<~0.1$~(GeV/$c)^2$~(23~$\pm$~4 and 21~$\pm$~4\%, respectively).\par

\indent (3)~The $\varepsilon^*_{ij}$ distributions also disagree with the functional form
$$
d\sigma/d\varepsilon^*~\simeq~\frac{1}{\pi}(1~+~c_1cos\varepsilon^*~+~c_2cos2\varepsilon^*),~~~~~(11)
$$
which follows from the statistical picture of the direct nuclear decay into the observed $\alpha$ particles. Assuming that in the c.m.s., each component of the $\alpha$ 3-momentum obeys the normal distribution $n(0,\sigma)$ (this automatically leads to the Rayleigh form of the distribution $d\sigma/dp^{*2}_T)$ and taking into account energy-momentum conservation in the decay, we find that the coefficients $c_1$ and $c_2$ in (11) are related to the azimuthal asymmetry $A^*$ and the azimuthal collinearity $B^*$ by the equations
$$
c_1~=~-(\pi/2)A^*~=~-(\pi/2)(N_{\alpha}~-~1)^{-1},
$$
$$
c_2~=~(\pi/2)B^*~=~(8\pi/25)(N_{\alpha}~-~1)^{-2},~~~~~(12)
$$
where (in our case) $N_{\alpha}$~=~4~[6]. The predicted forms (11) with coefficients (12) are represented by the curves in Fig. 5, and the values of $A^*$ and $B^*$ are given in the table. The discrepancy between the predictions and experimental data is clearly seen.\par

\indent (4)~In relation to the predictions based on transverse-momentum conservation, the experimental distribution $d\sigma/d\varepsilon^*_{ij}$ shows a higher degree of collinearity between the $\alpha$-particle transverse momenta $\mathbf p_T$ in the transverse plane (see the observed excess of experimental values over theoretical results for $\varepsilon_{ij}~\approx~0$ and $\varepsilon_{ij}~\approx~\pi$ in Fig. 5; the values of $B^*$ are given in the table).\par

\begin{table}
\caption{\label{}}
\begin{tabular}{l|c|c|c|c|c|c}
\hline\noalign{\smallskip}   
& \multicolumn{6}{c}{Quantity} \\ 
Event sample & $\langle p^*_T\rangle$, & $\langle p^{*2}_T\rangle^{1/2}$, & $A^*$ & $B^*$ & $\alpha^*$ & $\beta^*$\\
			 & MeV/$c$ & VeV/$c$ & & & & \\
\hline\noalign{\smallskip}
Experiment, all events&$121\pm2$&$145\pm3$&$-0.28\pm0.02$&$0.27\pm0.02$&$-0.23\pm0.01$&$0.22\pm0.02$\\
Experiment, &$115\pm2$&$134\pm4$&$-0.27\pm0.02$&$0.30\pm0.02$&$-0.23\pm0.01$&$0.25\pm0.02$\\
events with $t'<0.1$~(MeV/$c)^2$&&&&&&\\
Statistical model, $^{16}$O$\rightarrow4\alpha$&120&-&-0.33&0.07&-0.26&0.06\\
Statistical model, &120&-&-0.33&0.11&-0.27&0.09\\
$^{16}$O$\rightarrow^8$Be+$\alpha+\alpha\rightarrow4\alpha$&&&&&&\\
Statistical model, &120&-&-0.33&0.10&-0.27&0.08\\
$^{16}$O$\rightarrow^8$Be+$^8$Be$\rightarrow4\alpha$&&&&&&\\
Statistical model, &120&-&-0.33&0.07&-0.27&0.06\\
$^{16}$O$\rightarrow^{12}$C$^*+\alpha\rightarrow4\alpha$&&&&&&\\
Statistical model, &120&-&-0.33&0.08&-0.27&0.06\\
$^{16}$O$\rightarrow^{12}$C$^*+\alpha\rightarrow^8$Be+$\alpha+\alpha\rightarrow4\alpha$&&&&&&\\
\hline\noalign{\smallskip}
\end{tabular}
\end{table}

\indent Let us now discuss the results. We will use the Feshbach-Huang-Goldhaber statistical theory of prompt fragmentation \cite{Feshbach,Goldhaber}. In this theory (which is widely employed in analyses of fragmentation of excited nuclei), the measured values of $\langle p^{*2}_T\rangle$ can be related to the temperature of the oxygen nucleus decaying through the channel (2). In energy units, the temperature is estimated as
$$
KT~=~\frac{A}{A-1}(\sigma^2_N/m_N),~~~~~(13)
$$
where $m_N$ is the nucleon mass, and $\sigma^2_N$ is the variance of the momentum distribution of intranuclear nucleons; the latter governs the variance of the momentum distribution for arbitrary fragments through the so-called parabolic law
$$
\sigma^2_F~=~\sigma^2_NA_F(A-A_F)/(A-1).~~~~~(14)
$$
Here, $A$ and $A_F$ are the mass numbers of the nucleus undergoing fragmentation and the fragment under consideration, respectively. In our case, we have $A~=~16$, $A_F~\equiv~A_{\alpha}~=~4$, and $\sigma^2_{\alpha}~=~\langle p^{*2}_T\rangle/2$. Using the experimental values of $\langle p^{*2}_T\rangle$ (see table) for the complete sample of events (2), we obtain $kT~=~3.7$~MeV. For the events with $葉'~<~0.1$~(GeV/$c)^2$ and with $葉'~>~0.1$~(GeV/$c)^2$, we arrive at $kT~=~3.2$ and 4.8 MeV, respectively.\par

\indent The above estimates indicate that the coherent dissociation $^{16}$O~$\rightarrow~4\alpha$ is characterized by a lower value of $kT~(kT~\simeq~3~-~3.5$~MeV) than conventional multifragmentation in reactions of the type $A~+~B~\rightarrow~\alpha~+~X$ that are investigated in inclusive experiments (see, for example, \cite{Bengus,Bhanjo,Bondarenko2,Adamovich}). The last statement remains valid even if we consider that in many experimental studies, the decay temperature was overestimated because of erroneously using the laboratory momentum characteristics of $\alpha$ particles without taking into account the transverse motion of the system undergoing fragmentation. The value that we obtained for $kT$ is also significantly lower than the nucleon binding energy in the oxygen nucleus. On the other hand, such a low value of temperature corresponds to a small energy-momentum transfer to the nucleus undergoing fragmentation, which is typical of coherent processes. If nonstatistical tails in the distributions shown in Fig. 4 are due to the decays of massive intermediate states (the cascade decays of $^{16}$O are discussed below), the true $kT$ value is even lower than that quoted above.\par

\indent Let us consider this issue in more detail. First, we address the question of whether there is $\mathbf p_T$ collinearity in individual events. For this purpose, we supplemented the analysis of the inclusive characteristics of azimuthal asymmetry and collinearity ($A^*$ and $B^*$) with the calculation of the asymmetry $\alpha^*$ and collinearity $\beta^*$ in individual events by the formulas \cite{Bondarenko}
$$
\alpha^*~=~\langle cos\varepsilon^*_{ij}\rangle~=~\sum_{i,j=1}^{4}cos\varepsilon^*_{ij}/N_{\alpha}(N_{\alpha}-1) (i~\neq~j),~~~~~(15)
$$
$$
\alpha^*~=~\langle cos2\varepsilon^*_{ij}\rangle~=~\sum_{i,j=1}^{4}cos2\varepsilon^*_{ij}/N_{\alpha}(N_{\alpha}-1) (i~\neq~j),~~~~~(16)
$$
$(-(N-1)^{-1}=-(1/3)\le\alpha^*, \beta^*\le1)$. Assuming that the components of $\mathbf p^*_T$ are normally distributed and that the transverse momentum is conserved in the decay, we find that the mean values $\langle\alpha^*\rangle$ and $\langle\beta^*\rangle$ of random variables $\alpha^*$ and $\beta^*$ are given by \cite{Bondarenko}
$$
\langle\alpha^*\rangle~=~-(\pi/4)(N_{\alpha}-1)^{-1}~=~-0.26,
$$
$$
\langle\beta^*\rangle~=~(4\pi/25)(N_{\alpha}-1)^{-2}~=~0.056.~~~~~(17)
$$
\par

\indent The distribution in the collinearity $\beta^*$ is plotted in Fig. 6 for the 428 $^{16}$O~$\rightarrow~4\alpha$ events with $-t'~<~0.1$~(GeV/$c)^2$. The observed mean values $\langle\alpha^*\rangle$ and $\langle\beta^*\rangle$ are presented in the last two columns of the table.
\par

\begin{figure}
    \includegraphics[width=3in]{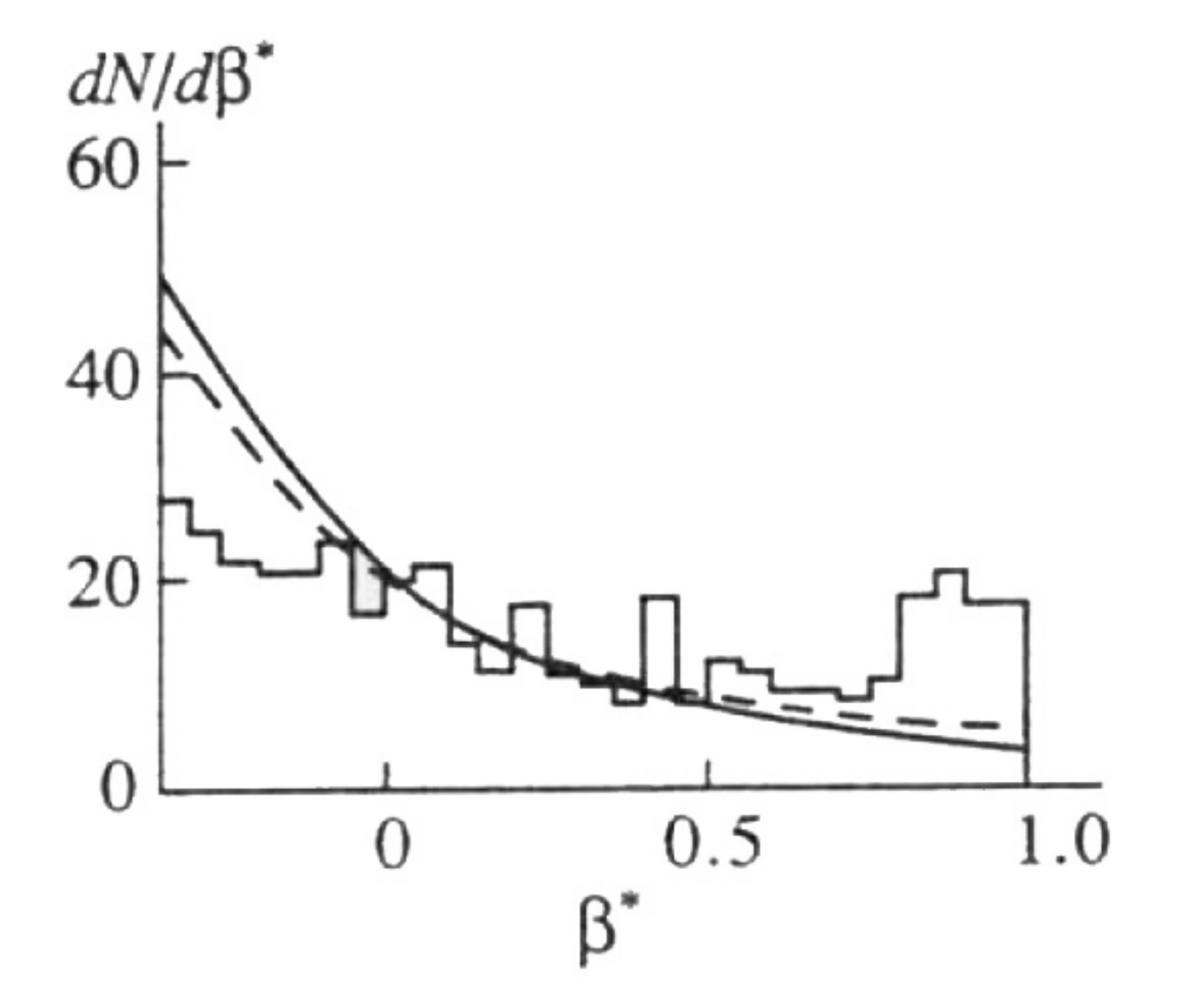}
    \caption{\label{Fig:6} Distribution in the azimuthal collinearity $\beta^*$ for an individual event. The solid and dashed curves represent the results of calculations made in the statistical theory for reacｬtion channels (18) and (19). respectively.}
    \end{figure}

\indent To evaluate the contributions of various $^{16}$O cascade decays to the observed tendency of the product $\alpha$ particles to diverge collinearly in the transverse plane, we performed a simulation of all possible cascade channels of dissociation according to the simplest statistical model \cite{Goldhaber}. Thus, we considered the transitions
$$
^{16}O~\rightarrow~4\alpha~(the~~direct~~channel),~~~~~(18)
$$
$$
^{16}O~\rightarrow~^8Be~+~\alpha~+~\alpha~\rightarrow~4\alpha,~~~~~(19)
$$
$$
^{16}O~\rightarrow~^8Be~+~^8Be~\rightarrow~4\alpha,~~~~~(20)
$$
$$
^{16}O~\rightarrow~^{12}C~+~\alpha~\rightarrow~4\alpha,~~~~~(21)
$$
$$
^{16}O~\rightarrow~^{12}C^+~\alpha~\rightarrow~^8Be~+~\alpha~+~\alpha~\rightarrow~4\alpha.~~~~~(22)
$$
For each decay fragment from reactions (18)-(22), we assumed that each 3-momentum component defined in the rest frame of the decaying nucleus or of the unstable intermediates state obeys a normal distribution $n(0,~\sigma^2)$, the variance $\sigma^2$ being dependent on the fragment mass according to the parabolic law (14). After the transformation from the rest frames of intermediate states to the c.m.s. of the $^{16}$O nucleus, we were left with the single model parameter $\sigma^2_N$ [see (14)]. This parameter was determined by imposing the requirement that the mean transverse momentum $\langle p^{*2}_T\rangle$ for the final-state $\alpha$ particles in each channel (18)-(22) be equal to its measured value~$\approx$~120~MeV/$c$ (see table). For each of the reactions (18)-(22) simulated according to the Monte Carlo method, the values of $A^*$, $B^*$, $\langle\alpha^*\rangle$, and $\langle\beta^*\rangle$ are shown in the table. The computed distributions $dN/d\beta^*$ are illustrated in Fig. 6 for channels (18) (the direct reaction) and (19) (under the above assumptions, the latter leads to the maximum collinearity of the vectors $\mathbf p^*_T$ in the final state).\par

\indent The results shown in the table and in Fig. 6 demonstrate that the observed effect of $\mathbf p^*_T$ collinearity cannot be attributed entirely to the cascade decays of $^{16}$O into $\alpha$ particles. If we go beyond the model from \cite{Goldhaber} by assuming that the temperature of primary $^{16}$O decays is substantially higher than that of the decays of intermediate fragments, a high degree of collinearity of the $\alpha$-particle transverse momenta in the final state can be obtained. In this case, however, theoretical calculations fail to describe other experimental characteristics (mean values $\langle p^*_T\rangle$, azimuthal asymmetries, etc.) under any assumptions about the relative contributions of channels (18) - (22).
\par

\indent Therefore, the only plausible way to explain the high values of $B^*$ and $\langle\beta^*\rangle$ is to assume that a nonzero angular momentum is transferred to the $^{16}$O nucleus in the collision. Needless to say, a comprehensive analysis of conclusions that can be drawn on basis of this assumption is beyond the scope of the present study. We look forward to testing it with increased statistics that will be accumulated in the near future.\par

\begin{acknowledgments}
\indent We are grateful to the administration and staff of the High Energy Division of JINR for the aid in performing our experiment at the synchrophasotron. We also acknowledge the efforts of all operators and technicians involved in scanning and measurements.\par
\end{acknowledgments}

\end{document}